\documentclass[11pt,a4paper]{article}

\usepackage[top=12mm,bottom=12mm,left=30mm,right=30mm,head=12mm,includeheadfoot]{geometry}
\usepackage{tocloft}

\usepackage[nottoc,notlot,notlof]{tocbibind}
\usepackage{titlesec}
\titlespacing*{\section}{0pt}{1.8\baselineskip}{\baselineskip}

\usepackage[utf8]{inputenc}
\usepackage{doi}
\usepackage{amsmath}
\usepackage{xcolor}
\usepackage{graphicx}
\usepackage{cite}
\usepackage[width=.90\textwidth]{caption}
\usepackage{hyperref}

\hypersetup{
    colorlinks,
    linkcolor={red!50!black},
    citecolor={blue!50!black},
    urlcolor={blue!80!black}
}

\usepackage[bitstream-charter]{mathdesign}
\DeclareSymbolFont{usualmathcal}{OMS}{cmsy}{m}{n}
\DeclareSymbolFontAlphabet{\mathcal}{usualmathcal}

\usepackage[most]{tcolorbox}
\usepackage{listings}
\usepackage{tikz}
\usepackage{xurl}
\usepackage{float}
\usetikzlibrary{arrows.meta,positioning}

\definecolor{codebg}{HTML}{0E1F20}
\definecolor{codepanel}{HTML}{0F4A4D}
\definecolor{codeteal}{HTML}{11A894}
\definecolor{codemint}{HTML}{88C9AA}
\definecolor{codeyellow}{HTML}{FED44E}
\definecolor{codered}{HTML}{D35E54}
\definecolor{codegray}{HTML}{B8DBC4}
\definecolor{codeorange}{HTML}{F9A25B}
\definecolor{codecream}{HTML}{FAEDBA}

\lstdefinestyle{dampfcode}{
    basicstyle=\ttfamily\small\color{codecream},
    keywordstyle=\color{codered},
    stringstyle=\color{codeyellow},
    commentstyle=\color{codegray},
    showstringspaces=false,
    breaklines=true,
    columns=fullflexible,
    keepspaces=true,
    frame=none
}

\tcbset{
    dampffigure/.style={
        enhanced,
        colback=codebg,
        boxrule=0pt,
        frame hidden,
        sharp corners=all,
        left=2mm,
        right=2mm,
        top=2mm,
        bottom=2mm,
        width=\linewidth,
        fonttitle=\ttfamily\small\color{codecream},
        coltitle=codecream,
        halign title=center,
        attach boxed title to top center={yshift=-1mm},
        boxed title style={
            colback=codepanel,
            boxrule=0pt,
            frame hidden,
            sharp corners=all
        }
    },
    dampfterminal/.style={
        enhanced,
        colback=codepanel,
        boxrule=0pt,
        frame hidden,
        sharp corners=all,
        left=1mm,
        right=1mm,
        top=1mm,
        bottom=1mm,
        width=\linewidth
    }
}

\begin{document}

\pagestyle{plain}

\begin{center}{\Large \textbf{
DAMPyF: a Python implementation of the DAMPF method for the simulation of open-system dynamics\\
}}\end{center}

\begin{center}\textbf{
N. Lorenzoni\textsuperscript{1$\star$},
S. F. Huelga\textsuperscript{1$\dagger$} and
M. B. Plenio\textsuperscript{1$\dagger\dagger$}
}\end{center}

\begin{center}
{\bf 1} Institute of Theoretical Physics, Ulm University, Albert-Einstein-Allee 11, 89081 Ulm, Germany
\\[\baselineskip]
$\star$ \href{mailto:nicola.lorenzoni@uni-ulm.de}{\small nicola.lorenzoni@uni-ulm.de}
\\
$\dagger$ \href{mailto:susana.huelga@uni-ulm.de}{\small susana.huelga@uni-ulm.de}
\\
$\dagger\dagger$ \href{mailto:martin.plenio@uni-ulm.de}{\small martin.plenio@uni-ulm.de}
\end{center}

\section*{Abstract}
\textbf{\boldmath{%
DAMPyF is an open-source Python implementation of the dissipation-assisted matrix product factorization (DAMPF) method, a tensor-network-based approach for the numerically exact simulation of finite-dimensional quantum systems coupled to bosonic environments. The method relies on a pseudomode representation of structured reservoirs and a matrix-product-state representation of the density matrix of the extended system, comprising the system and the pseudomodes. DAMPyF currently provides two workflows. First, it supports excitation energy-transfer dynamics within the single-system-excitation manifold, in which a system excitation is propagated in time. Second, it provides a high-level workflow tailored to molecular spectroscopy, in which the system levels represent electronic states and optical coherences are propagated for the subsequent computation of linear spectra, including absorption and circular dichroism. This paper describes the physical model, the DAMPF algorithm, the user-facing code structure, installation and execution, input and output formats, and minimal examples.
}}

\vspace{\baselineskip}

\vspace{10pt}
\noindent\rule{\textwidth}{1pt}
\tableofcontents
\noindent\rule{\textwidth}{1pt}
\vspace{10pt}

\section{Introduction}
\label{sec:introduction}

Open quantum systems provide a natural framework for describing situations in which a finite set of degrees of freedom, referred to as the system, evolves under the influence of a surrounding environment~\cite{Rivas2011, BreuerPetruccioneBook, VacchiniOpenQuantumSystems2024}. In many applications, the system is not weakly coupled to a featureless bath but, in sharp contrast, real world systems frequently exhibit memory effects that may persist on the timescale of the system dynamics, and system--environment correlations that can play an essential role~\cite{RivasRPG2014, BreuerRMP2016, deVegaRevModPhys2017}; a regime referred to as \emph{non-perturbative} and comprising a wide variety of physical processes.

This is, for example, the case for electronic dynamics in molecular aggregates, where excitation-energy-transfer dynamics, charge-transfer dynamics, and optical spectra are shaped by the coupling between electronic states and vibrational environments~\cite{CaycedoNatCommun2022,MayKuhnBook}. The spectral densities characterizing these vibrational environments are typically highly structured in realistic systems, and their simulation therefore requires methods capable of representing several electronic sites and many environmental modes while keeping the computational cost manageable.

Similar classes of non-perturbative models arise for quantum emitters interacting with structured photonic environments, where environmental memory can strongly modify emission and excitation-transfer dynamics~\cite{PriorPRA2013, MedinaPRL2021,SanchezBarquillaNanophotonics2022}, as well as in trapped-ion quantum simulators of spin--boson models, where structured bosonic environments can be engineered and controlled~\cite{Lemmer_2018, SoSciAdv2024, SunNC2025}.

In this non-perturbative regime, growing attention has been drawn toward \emph{numerically-exact} methods, namely methods that do not rely on uncontrolled approximations and can solve the equations of motion with arbitrarily high accuracy~\cite{RMP}.

Tensor-network methods address this challenge by exploiting compressed representations of many-body quantum states and operators~\cite{SchollwockAnnPhys2011,OrusAnnPhys2014,OrusNatRev2019}. The dissipation-assisted matrix product factorization (DAMPF) method combines this idea with a pseudomode representation of structured bosonic environments~\cite{SomozaPRL2019}. The pseudomode approach, which falls under the umbrella of Markovian-embedding constructions, has been widely employed in simulations of bosonic environments~\cite{ImamogluPRA1994,GarrawayPRA1997,DaltonPRA2001,MartinazzoJChemPhys2011, MenczelPRR2024}. In this approach, the environment is replaced by a finite set of damped harmonic modes, referred to as pseudomodes, chosen to reproduce the relevant bath correlation functions. Under the corresponding assumptions, this condition preserves the reduced system dynamics and enables numerically exact techniques~\cite{TamascelliPRL2018}.

\begin{figure}[ht]
    \centering
    \includegraphics[width=\linewidth]{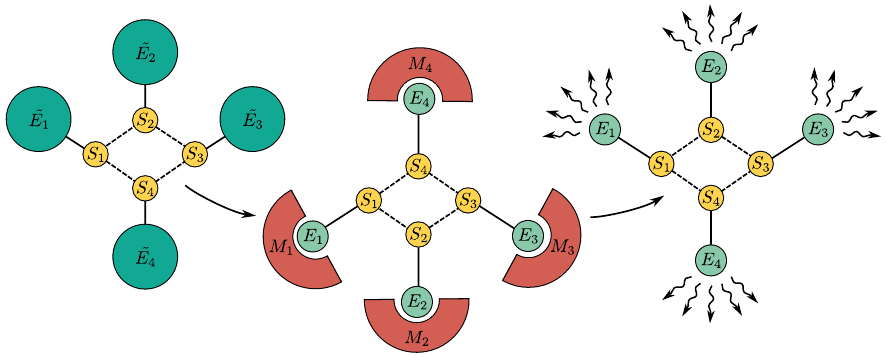}
    \caption{Conceptual representation of the pseudomode approach. The physical local environments ($\tilde{E}_i$) associated with each site ($S_i$) are partitioned into a non-Markovian core ($E_i$) and a Markovian remainder ($M_i$), which is traced out exactly and induces a Lindblad dissipator.}
    \label{Fig1}
\end{figure}

The pseudomode approach effectively partitions the bosonic environment into two components: a non-Markovian core retaining memory effects and a Markovian remainder that can be mathematically traced out, inducing a computationally tractable Lindblad dissipator, as represented schematically in Fig.~\ref{Fig1}. A continuous, highly structured environment can therefore be represented using a modest number of pseudomodes. Furthermore, their local dissipation suppresses the growth of correlations in the tensor-network representation, making simulations with many environmental modes feasible~\cite{SomozaPRL2019,SomozaCommunPhys2023, LorenzoniPRL2024,LorenzoniSciAdv2025}.

DAMPyF is a Python implementation of the DAMPF method. It is intended to provide a readable and reproducible codebase for simulations of multi-site systems in which each site may couple to local pseudomode environments. The code is organized around a single user-editable configuration file and a small set of text input files for system Hamiltonians, pseudomode parameters, and optional electric and magnetic transition dipoles for the computation of linear spectra in molecular systems. The current release provides two user-facing simulation modes: energy-transfer dynamics and linear-spectra propagation.

The paper is organized as follows. Section~\ref{sec:method} first summarizes the physical model and equation of motion implemented in DAMPyF, allowing readers to assess whether the code is suitable for their problem. The same section then describes the DAMPF ansatz and time-evolution strategy for readers interested in the internal algorithm. Sections~\ref{sec:workflow}--\ref{sec:examples} form the user-oriented part of the paper: they describe the repository workflow, installation and execution, configuration variables and input files, output and post-processing, convergence parameters, and minimal examples. Appendix~\ref{app:pseudomode_parameters} describes the characterization of pseudomode parameters, while Appendix~\ref{app_spectra} discusses the theory and simulation of linear spectra.

\section{The DAMPF method: model and algorithm}
\label{sec:method}

\begin{figure}[ht]
    \centering
    \includegraphics[width=\linewidth]{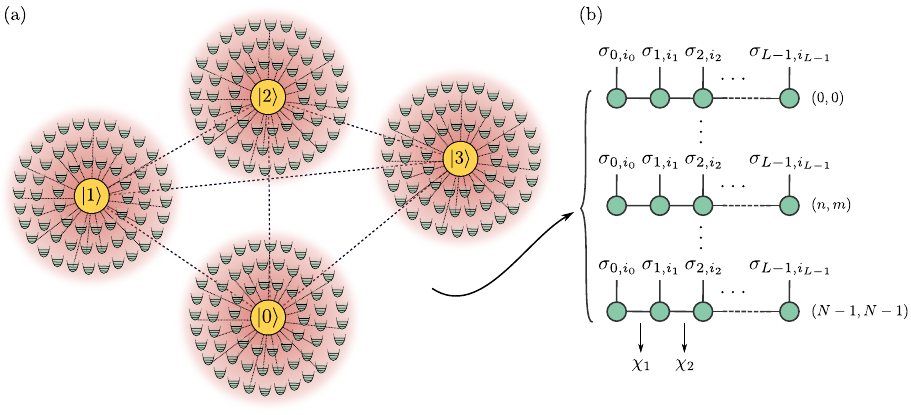}
    \caption{Graphical representation of (a) the extended system, comprising the finite-dimensional system and its local pseudomode environments, and (b) the DAMPF decomposition of the extended density matrix into system-indexed pseudomode MPSs.}
    \label{Fig2}
\end{figure}

This section first summarizes the physical model and equation of motion central to DAMPF, thereby clarifying the class of systems addressed by the approach. It then presents the internal algorithm used for the DAMPF propagation. This second part is not required for running the predefined workflows of the software package, but is relevant for readers interested in the details of the implementation.

\subsection{Physical model}
\label{sec:physical_model}

The DAMPF method is formulated to simulate the dynamics of generalized spin--boson models, where a finite set of system sites is coupled to local pseudomode environments~\cite{MayKuhnBook}.

Typically, DAMPF considers the dynamics within the single-excitation manifold of the system. The Hamiltonian of the extended system, given by the system plus the pseudomodes, is written as
\begin{equation}
\label{eq:hamiltonian}
H=H_{\mathrm{s}}+H_{\mathrm{e}}+H_{\mathrm{s-e}}.
\end{equation}
The system Hamiltonian is 
\begin{equation}
\label{eq:hamiltonian_system}
H_{\mathrm{s}} = \sum_{n=0}^{N-1}\epsilon_n\left|n\right\rangle\!\left\langle n\right|+
\sum_{0\leq m<n\leq N-1}\left(V_{mn}\left|m\right\rangle\!\left\langle n\right|+\mathrm{h.c.}\right),
\end{equation}
where $\left|n\right\rangle$ labels a local excitation of site $n$ with energy $\epsilon_n$ and $V_{mn}$ is the inter-site coupling. The pseudomode Hamiltonian is
\begin{equation}
\label{eq:hamiltonian_pseudomodes}
H_{\mathrm{e}} = \sum_{n=0}^{N-1}\sum_{q=0}^{Q_n-1}\omega_{nq}a_{nq}^{\dagger}a_{nq},
\end{equation}
where $a^\dagger_{nq}$ and $a_{nq}$ are, respectively, the creation and annihilation operators associated with the $q$-th pseudomode coupled to site $n$. Finally, the system--pseudomode coupling Hamiltonian is
\begin{equation}
\label{eq:hamiltonian_coupling}
H_{\mathrm{s-e}} = \sum_{n=0}^{N-1}\left|n\right\rangle\!\left\langle n\right|\otimes
\sum_{q=0}^{Q_n-1}\left(g_{nq}a_{nq}^{\dagger}+g_{nq}^*a_{nq}\right),
\end{equation}
where $g_{nq}$ is the system--pseudomode coupling strength. The model is represented schematically in Fig.~\ref{Fig2}(a).

DAMPF simulates the density matrix of the extended system, $\rho_{\mathrm{se}}(t)$, by solving the Gorini--Kossakowski--Sudarshan--Lindblad (GKSL) master equation~\cite{GoriniJMathPhys1976,LindbladCommunMathPhys1976}
\begin{equation}
\label{eq:gksl}
\frac{\mathrm{d}}{\mathrm{d}t}\rho_{\mathrm{se}}(t)
=
-\mathrm{i}[H,\rho_{\mathrm{se}}(t)]
+\mathcal{D}_{\mathrm{e}}[\rho_{\mathrm{se}}(t)]
\equiv
\mathcal{L}\rho_{\mathrm{se}}(t),
\end{equation}
where $\mathcal{L}$ is the Liouvillian and $\mathcal{D}_{\mathrm{e}}$ represents the Lindblad relaxation of the pseudomodes.
Equations~\eqref{eq:hamiltonian}--\eqref{eq:gksl} are written in units where the factor \(\hbar\) does not appear explicitly.
The dissipator is given by
\begin{equation}
\label{eq:dissipator}
\begin{aligned}
\mathcal{D}_{\mathrm{e}}[\rho_{\mathrm{se}}]
&=
\sum_{n=0}^{N-1}\sum_{q=0}^{Q_n-1}
\gamma_{nq}
\Bigg[
\left(1+\bar n_{nq}\right)
\left(
a_{nq}\rho_{\mathrm{se}} a_{nq}^{\dagger}
-\frac{1}{2}\{a_{nq}^{\dagger}a_{nq},\rho_{\mathrm{se}}\}
\right)
\\
&\hspace{3.0cm}
+
\bar n_{nq}
\left(
a_{nq}^{\dagger}\rho_{\mathrm{se}} a_{nq}
-\frac{1}{2}\{a_{nq}a_{nq}^{\dagger},\rho_{\mathrm{se}}\}
\right)
\Bigg],
\end{aligned}
\end{equation}
where $\gamma_{nq}$ is the damping rate and
\begin{equation}
\bar n_{nq}=\left[\exp\left(\frac{\omega_{nq}}{k_{\mathrm{B}}T_{nq}}\right)-1\right]^{-1}
\end{equation}
is the thermal occupation of pseudomode $(n,q)$ for $k_{\mathrm B}T_{nq}>0$. For zero thermal energy, DAMPyF sets $\bar n_{nq}=0$. Negative-frequency pseudomodes are permitted only at zero thermal energy and therefore also have $\bar n_{nq}=0$. In this formulation, each positive-frequency pseudomode may in general be assigned its own temperature. For details regarding the characterization of the pseudomode parameters appearing in $H_{\mathrm{e}}$, $H_{\mathrm{s-e}}$ and $\mathcal{D}_{\mathrm{e}}$, refer to Appendix~\ref{app:pseudomode_parameters}.

For a fresh simulation, the initial state in DAMPF is taken to be factorized,
\begin{equation}
\label{eq:initial_state}
\rho_{\mathrm{se}}(0)=\rho_{\mathrm{s}}(0)\otimes\bigotimes_{n,q}\rho_{\mathrm{th}}^{nq},
\end{equation}
where $\rho_{\mathrm{s}}(0)$ is the initial state of the system, and $\rho_{\mathrm{th}}^{nq}$ is the thermal state of pseudomode $(n,q)$.
System observables are obtained from the reduced system density matrix, which is computed by tracing out the pseudomode degrees of freedom,
\begin{equation}
\label{eq:reduced_system_density_matrix}
\rho_{\mathrm{s}}(t)
=
\operatorname{Tr}_{\mathrm{e}}\!\left[\rho_{\mathrm{se}}(t)\right].
\end{equation}

When a simulation is resumed from a checkpoint, the loaded MPS may instead contain system--pseudomode correlations generated during the preceding propagation. In its current implementation, DAMPyF applies the DAMPF method within the single-excitation manifold of the system. The preceding equations define the physical class of problems addressed by DAMPyF. A reader interested only in preparing and running standard simulations can proceed from here to Sec.~\ref{sec:workflow}. The following subsections describe the tensor-network representation and propagation algorithm used internally by the code.

\subsection{DAMPF ansatz}
\label{sec:dampf_ansatz}

In DAMPF, the density matrix of the extended system is first decomposed with respect to the system degrees of freedom as
\begin{equation}
\label{eq:dampf_ansatz}
\rho_{\mathrm{se}}(t)
=
\sum_{m,n}
\left|m\right\rangle\!\left\langle n\right|
\otimes
\rho_{\mathrm{e}}^{(m,n)}(t).
\end{equation}
That is, the system density-matrix indices $(m,n)$ are kept explicit, with each system matrix element associated with a pseudomode operator $\rho_{\mathrm{e}}^{(m,n)}(t)$. These pseudomode operators are then represented using the matrix-product-state (MPS) formalism.

Let the full set of pseudomodes be labeled by the index $\ell=0,\dots,L-1$, with $L=\sum_{n=0}^{N-1}Q_n$. For each pseudomode $\ell$, we introduce a local---that is, single-pseudomode---operator basis $\{\sigma_{\ell,i_\ell}\}_{i_\ell=0}^{d_\ell^2-1}$, where $d_\ell$ is the number of retained levels for that pseudomode. The pseudomode operator associated with the system matrix element $(m,n)$ is then written in MPS form as
\begin{equation}
\label{eq:vib_mps}
\rho_{\mathrm{e}}^{(m,n)}
=
\sum_{i_0,\ldots,i_{L-1}}
A_{0;i_0}^{(m,n)}
A_{1;i_1}^{(m,n)}
\cdots
A_{L-1;i_{L-1}}^{(m,n)}
\,
\sigma_{0,i_0}\otimes\sigma_{1,i_1}\otimes\cdots\otimes\sigma_{L-1,i_{L-1}}.
\end{equation}
Here, each tensor has components
\begin{equation}
A_{\ell;i_\ell}^{(m,n)}
\equiv
A_{\ell;\alpha_{\ell},i_\ell,\alpha_{\ell+1}}^{(m,n)},
\end{equation}
where $i_\ell=0,\dots,d_\ell^2-1$ is the local physical index associated with the operator basis of pseudomode $\ell$, while $\alpha_\ell$ and $\alpha_{\ell+1}$ are virtual indices of dimensions $\chi_\ell$ and $\chi_{\ell+1}$, referred to as bond dimensions. This is represented graphically in Fig.~\ref{Fig2}(b).

According to this construction, each system operator $\left|m\right\rangle\!\left\langle n\right|$ is associated with a different MPS, labeled by the pair $(m,n)$. Conceptually, the DAMPF ansatz therefore contains $N^2$ MPS components, one for each system density-matrix element. In energy-transfer simulations, DAMPyF exploits Hermiticity and stores only the $N(N+1)/2$ diagonal and upper-triangular components, reconstructing the lower-triangular entries by conjugation when observables are evaluated. In linear-spectra simulations, all $N^2$ optical-coherence components are stored explicitly.

The local operator basis implemented in DAMPyF is Hilbert--Schmidt orthonormal. It consists of the normalized identity $\mathbb{I}/\sqrt{d_\ell}$ together with $d_\ell^2-1$ traceless matrices. This basis is particularly convenient for trace operations in the MPS formalism.

\subsection{On the rationale behind the DAMPF ansatz}

This representation is efficient in regimes where the number of relevant system states grows polynomially with the system size $N$. It differs from a single-MPS representation of the full system-plus-pseudomode density matrix, which would be the natural choice when the system Hilbert space itself grows exponentially with $N$. By keeping the system indices outside the MPS representation, the MPSs are defined only over the pseudomode degrees of freedom. As a consequence, the correlations encoded in the bond dimension of each individual MPS describe correlations among pseudomodes, rather than direct system--pseudomode correlations.

Since the pseudomodes are not directly coupled to one another, these correlations are typically weaker than the direct correlations between the system and the pseudomodes that would appear in a full MPS representation. As a result, the DAMPF representation can often use smaller MPS bond dimensions than a full MPS representation of all degrees of freedom. Moreover, the local dissipation of the pseudomodes remains local at the level of the tensor-network representation. This local damping further suppresses the growth of correlations within the MPSs and contributes to the efficiency of the DAMPF ansatz~\cite{SomozaPRL2019}.

\subsection{Time evolution}
\label{sec:time_evolution}

The equation of motion in Eq.~\eqref{eq:gksl} can equivalently be written in integrated form as
\begin{equation}
\rho(t)=e^{\mathcal{L}t}\rho(0).
\end{equation}
Directly evaluating this expression becomes computationally costly when many degrees of freedom are retained. This complexity can be addressed through decomposition strategies.

Introducing a finite time step $dt$ and considering discretized times $t_k=k\,dt$, with $k=0,1,2,\ldots$, one obtains
\begin{equation}
\rho(t_k)=\left(e^{\mathcal{L}dt}\right)^k\rho(0).
\end{equation}
The problem is therefore reduced to the repeated application of the short-time propagator $e^{\mathcal{L}dt}$.

This can be simplified by decomposing the Liouvillian in Eq.~\eqref{eq:gksl} into system, pseudomode, system--pseudomode coupling, and pseudomode relaxation contributions, namely
\begin{equation}
\label{eq:liouvillian_decomposition}
\mathcal{L}=\mathcal{L}_{\mathrm{s}}+\mathcal{L}_{\mathrm{e}}+\mathcal{L}_{\mathrm{s-e}}+\mathcal{D}_{\mathrm{e}}.
\end{equation}
The short-time propagator can then be approximated using a first-order Trotter decomposition~\cite{TrotterPAMS1959},
\begin{equation}
\label{eq:trotter}
e^{\mathcal{L}dt}
=
e^{\mathcal{L}_{\mathrm{s}}dt}
e^{\mathcal{L}_{\mathrm{e}}dt}
e^{\mathcal{L}_{\mathrm{s-e}}dt}
e^{\mathcal{D}_{\mathrm{e}}dt}
+\mathcal{O}(dt^2),
\end{equation}

thereby reducing it to a product of elementary propagators associated with the individual contributions to the Liouvillian.

The error induced by this decomposition can be reduced using higher-order product formulas~\cite{suzuki_generalized_1976}. For example, a symmetric second-order Trotter--Suzuki decomposition gives~\cite{Suzuki1985}
\begin{equation}
\label{eq:symmtrotter}
\begin{aligned}
e^{\mathcal{L}dt}
&=
e^{\mathcal{D}_{\mathrm{e}}dt/2}
e^{\mathcal{L}_{\mathrm{s-e}}dt/2}
e^{\mathcal{L}_{\mathrm{e}}dt/2}
e^{\mathcal{L}_{\mathrm{s}}dt}
e^{\mathcal{L}_{\mathrm{e}}dt/2}
e^{\mathcal{L}_{\mathrm{s-e}}dt/2}
e^{\mathcal{D}_{\mathrm{e}}dt/2}
+\mathcal{O}(dt^3),
\end{aligned}
\end{equation}

which employs the same superoperators as the first-order Trotter decomposition, with rescaled time steps.

In terms of the Hamiltonian in Eq.~\eqref{eq:hamiltonian}, elementary propagators read

\begin{equation}
\begin{aligned}
\label{eq:elementary_propagators_general}
e^{\mathcal{L}_{\mathrm{s}}dt}[\rho_{\mathrm{se}}]
&=
e^{-\mathrm{i}H_{\mathrm{s}}dt}\rho_{\mathrm{se}}e^{\mathrm{i}H_{\mathrm{s}}dt},\\
e^{\mathcal{L}_{\mathrm{e}}dt}[\rho_{\mathrm{se}}]
&=
e^{-\mathrm{i}H_{\mathrm{e}}dt}\rho_{\mathrm{se}}e^{\mathrm{i}H_{\mathrm{e}}dt},\\
e^{\mathcal{L}_{\mathrm{s-e}}dt}[\rho_{\mathrm{se}}]
&=
e^{-\mathrm{i}H_{\mathrm{s-e}}dt}\rho_{\mathrm{se}}e^{\mathrm{i}H_{\mathrm{s-e}}dt}.
\end{aligned}
\end{equation}

\subsubsection{Time evolution in MPS form}
\label{sec:mps_time_evolution}

The elementary propagators have a simple structure when applied to the DAMPF ansatz in Eq.~\eqref{eq:dampf_ansatz}. As shown below, the pseudomode, system--pseudomode-coupling, and dissipative propagators act locally on the individual MPS tensors, whereas the system propagator combines the different MPSs labeled by the system indices $(m,n)$.

\paragraph{Pseudomode propagator.}

The free pseudomode Hamiltonian is a sum of single-pseudomode terms,
\begin{equation}
\label{eq:mps_local_pseudomode_hamiltonian}
H_{\mathrm{e}}
=
\sum_{\ell=0}^{L-1}H_{\mathrm{e}}^{\ell},
\end{equation}
where $\ell$ labels the pseudomodes. Since these terms act on different local Hilbert spaces, the free pseudomode propagator factorizes into local maps. Consequently, the free pseudomode propagator is represented in MPS form by a matrix-product operator (MPO) with bond dimension \(\chi_{\mathrm{MPO}}=1\), whose tensors are determined via
\begin{equation}
\label{eq:pseudomode_local_tensor_update}
A_{\ell;i'_\ell}^{(m,n)}(t_{k+1})
=
\sum_{i_\ell}
O_{\ell;i'_\ell i_\ell}^{[\mathrm{e}]}
A_{\ell;i_\ell}^{(m,n)}(t_k),
\end{equation}
with
\begin{equation}
\label{eq:pseudomode_local_tensor}
O_{\ell;i'_\ell i_\ell}^{[\mathrm{e}]}
=
\operatorname{Tr}
\left[
\sigma_{\ell,i'_\ell}^{\dagger}
e^{-\mathrm{i}H_{\mathrm{e}}^{\ell}dt}
\sigma_{\ell,i_\ell}
e^{\mathrm{i}H_{\mathrm{e}}^{\ell}dt}
\right].
\end{equation}

This update is the same for all system components $(m,n)$.

\paragraph{System--pseudomode coupling propagator.}

The system--pseudomode coupling Hamiltonian can be rewritten as
\begin{equation}
\label{eq:coupling_hamiltonian_local}
H_{\mathrm{s-e}}
=
\sum_{n=0}^{N-1}
|n\rangle\!\langle n|
\otimes
\sum_{q=0}^{Q_n-1}
B_{nq},
\qquad
B_{nq}
=
\left(g_{nq}
a_{nq}^{\dagger}+g_{nq}^*a_{nq}\right).
\end{equation}
It is therefore diagonal in the site basis. As a consequence, the coupling propagator does not mix different pairs $(m,n)$ and acts independently on each MPS $\rho_{\mathrm{e}}^{(m,n)}$.

For the pseudomode $q$ coupled to site $r$, the corresponding local MPS update is
\begin{equation}
\label{eq:coupling_local_tensor_update}
A_{rq;i'_{rq}}^{(m,n)}(t_{k+1})
=
\sum_{i_{rq}}
O_{rq;i'_{rq} i_{rq}}^{[\mathrm{s-e}](m,n)}
A_{rq;i_{rq}}^{(m,n)}(t_k),
\end{equation}
where
\begin{equation}
\label{eq:coupling_local_tensor_cases}
O_{rq;i'_{rq} i_{rq}}^{[\mathrm{s-e}](m,n)}
=
\begin{cases}
\delta_{i'_{rq} i_{rq}},
& r\neq m \ \text{and}\ r\neq n,\\[1ex]
\operatorname{Tr}
\left[
\sigma_{rq,i'_{rq}}^{\dagger}
e^{-\mathrm{i}B_{rq}dt}
\sigma_{rq,i_{rq}}
e^{\mathrm{i}B_{rq}dt}
\right],
& r=m=n,\\[1ex]
\operatorname{Tr}
\left[
\sigma_{rq,i'_{rq}}^{\dagger}
e^{-\mathrm{i}B_{rq}dt}
\sigma_{rq,i_{rq}}
\right],
& r=m\neq n,\\[1ex]
\operatorname{Tr}
\left[
\sigma_{rq,i'_{rq}}^{\dagger}
\sigma_{rq,i_{rq}}
e^{\mathrm{i}B_{rq}dt}
\right],
& r=n\neq m.
\end{cases}
\end{equation}
The coupling propagator is thus represented as a collection of MPOs with bond dimension $\chi_{\mathrm{MPO}}=1$, one for each component $(m,n)$.

\paragraph{Lindblad relaxation propagator.}

The Lindblad relaxation propagator is also local in the pseudomode degrees of freedom. The dissipator can be decomposed as
\begin{equation}
\label{eq:mps_local_dissipator_decomposition}
\mathcal{D}_{\mathrm{e}}
=
\sum_{\ell=0}^{L-1}
\mathcal{D}_{\mathrm{e}}^{\ell}.
\end{equation}
For each pseudomode, the local dissipative map can be represented in vectorized Liouville space. We use the row-vectorization convention, for which
\begin{equation}
\label{eq:row_vectorisation_convention}
A\rho B
\quad\longrightarrow\quad
(A\otimes B^{\mathsf{T}})\operatorname{vec}(\rho).
\end{equation}
With this convention, the local dissipator for pseudomode $\ell$ is represented by the matrix
\begin{equation}
\label{eq:local_dissipator_matrix}
\begin{aligned}
\mathbf{D}_{\ell}
&=
\gamma_{\ell}(1+\bar n_{\ell})
\left(
a_{\ell}\otimes a_{\ell}^{*}
-\frac{1}{2}a_{\ell}^{\dagger}a_{\ell}\otimes\mathbb{I}
-\frac{1}{2}\mathbb{I}\otimes a_{\ell}^{\mathsf{T}}a_{\ell}^{*}
\right)\\
&\quad+
\gamma_{\ell}\bar n_{\ell}
\left(
a_{\ell}^{\dagger}\otimes a_{\ell}^{\mathsf{T}}
-\frac{1}{2}a_{\ell}a_{\ell}^{\dagger}\otimes\mathbb{I}
-\frac{1}{2}\mathbb{I}\otimes a_{\ell}^{*}a_{\ell}^{\mathsf{T}}
\right).
\end{aligned}
\end{equation}
Denoting the vectorized representation of the local operator-basis element $\sigma_{\ell,i_\ell}$ by $\left|\sigma_{\ell,i_\ell}\right)$, the local MPS update reads
\begin{equation}
\label{eq:dissipative_local_tensor_update}
A_{\ell;i'_\ell}^{(m,n)}(t_{k+1})
=
\sum_{i_\ell}
O_{\ell;i'_\ell i_\ell}^{[\mathcal{D}]}
A_{\ell;i_\ell}^{(m,n)}(t_k),
\end{equation}
where
\begin{equation}
\label{eq:dissipative_local_tensor}
O_{\ell;i'_\ell i_\ell}^{[\mathcal{D}]}
=
\left(
\sigma_{\ell,i'_\ell}
\right|
\exp\left(\mathbf{D}_{\ell}dt\right)
\left|
\sigma_{\ell,i_\ell}
\right).
\end{equation}
The relaxation propagator is therefore also represented as an MPO with bond dimension \(\chi_{\mathrm{MPO}}=1\).

\paragraph{System propagator.}

The system propagator is generated by
\begin{equation}
\label{eq:system_propagator}
U_{\mathrm{s}}(dt)
=
\exp[-\mathrm{i}H_{\mathrm{s}}dt].
\end{equation}
Since the system Hamiltonian is in general not diagonal in the site basis due to the presence of the inter-site coupling, this propagator mixes the pseudomode operators according to
\begin{equation}
\begin{aligned}
\label{eq:system_mixing}
\rho_{\mathrm{e}}^{(m,n)}(t_{k+1})
&=
\sum_{m',n'}
U_{\mathrm{s},mm'}(dt)\,
U^{*}_{\mathrm{s},nn'}(dt)\,
\rho_{\mathrm{e}}^{(m',n')}(t_k)\\
&=\sum_{m'}
U_{\mathrm{s},mm'}(dt)\,
\Big(\sum_{n'}U^{*}_{\mathrm{s},nn'}(dt)\,
\rho_{\mathrm{e}}^{(m',n')}(t_k)\Big).
\end{aligned}
\end{equation}
Thus, each updated MPS labeled by $(m,n)$ is obtained as a linear combination of the MPSs describing the previous time step configuration.

Since the sum of MPSs generally produces an MPS whose bond dimension is the sum of the bond dimensions of the individual terms, DAMPyF combines the system-propagator update with compression.

\section{DAMPyF setup}
\label{sec:workflow}

The DAMPyF source code is available on \href{https://github.com/NicolaLorenzoni/DAMPyF}{GitHub}.

A typical DAMPyF calculation requires the user to provide the system and pseudomode input files and edit the main configuration file. The code can then be launched and the generated output analyzed.

The main user-facing files are organized as follows:

\begin{center}
\begin{tcolorbox}[dampffigure,title={DAMPyF repository workflow}]
\begin{minipage}{0.65\linewidth}
\begin{tcolorbox}[dampfterminal]
{\ttfamily\small\color{codecream}
\begin{tabular}{@{}l@{}}
{DAMPyF/}\\
|- \textcolor{codered}{configure\_simulation.py}\\
|- \textcolor{codeorange}{run\_dampf.py}\\
|- \textcolor{codeorange}{slurm\_code\_launcher.sh}\\
|- {dampf\_modules/}\\
|- \textcolor{codeyellow}{input\_data/}\\
|\hspace*{1.2em}|- \textcolor{codeyellow}{system/}\\
|\hspace*{1.2em}|- \textcolor{codeyellow}{pseudomodes/}\\
|- \textcolor{codeteal}{analysis\_tools/}\\
|\hspace*{1.2em}|- \textcolor{codeteal}{plot\_system\_dynamics.py}\\
|\hspace*{1.2em}|- \textcolor{codeteal}{compute\_absorption\_spectrum.py}\\
|\hspace*{1.2em}|- \textcolor{codeteal}{compute\_circular\_dichroism\_spectrum.py}\\
|- {helper\_tools/}\\
|- {output\_data/}
\end{tabular}
}
\end{tcolorbox}
\end{minipage}
\hfill
\begin{minipage}{0.34\linewidth}
\centering
{\ttfamily\small\color{codegray}User-facing workflow:}
\vspace{2mm}

\begin{tikzpicture}[
    node distance=0.45cm,
    box/.style={
        draw=none,
        fill=codepanel,
        align=center,
        text=codecream,
        font=\ttfamily\scriptsize,
        inner sep=5pt,
        minimum width=0.90\linewidth
    },
    arrow/.style={-{Latex[length=2mm]}, draw=codeteal, thick}
]
\node[box] (a) {Provide input parameters\\\textcolor{codeyellow}{input\_data/}};
\node[box, below=of a] (b) {Configure simulation\\\textcolor{codered}{configure\_simulation.py}};
\node[box, below=of b] (c) {Launch code\\\textcolor{codeorange}{run\_dampf.py}\\\textcolor{codeorange}{slurm\_code\_launcher.sh}};
\node[box, below=of c] (d) {Analyze output\\\textcolor{codeteal}{analysis\_tools/}};
\draw[arrow] (a) -- (b);
\draw[arrow] (b) -- (c);
\draw[arrow] (c) -- (d);
\end{tikzpicture}
\end{minipage}
\end{tcolorbox}
\end{center}

The file \path{configure_simulation.py} contains the user-facing simulation parameters, while \path{run_dampf.py} launches the propagation, with optional deployment on Slurm-based machines via the \path{slurm_code_launcher.sh} file. Input files are stored in \path{input_data/} and output files are written to \path{output_data/}.

Default post-processing scripts used to generate plots are collected in \path{analysis_tools/}, while utilities for preparing input data and estimating numerical requirements are provided in \path{helper_tools/}.

\subsection{Input data}
\label{sec:input_data}

The input files are plain-text files. The system parameters are stored in \path{input_data/system/}, together with the optional transition-dipole files used in linear-spectra simulations of molecular systems. In the linear-spectra workflow, the system represents an electronic system within the single-excitation manifold.

For a dimeric system with $N=2$, examples of the system input files are:

\begin{center}
\begin{tcolorbox}[dampffigure,title={System input files}]

{\ttfamily\small\color{codeyellow}input\_data/system/dimer\_system\_hamiltonian.txt}
\vspace{1mm}
\begin{tcolorbox}[dampfterminal]
\begin{lstlisting}[style=dampfcode]
# system Hamiltonian in site basis
1000.0   100.0
 100.0  1200.0
\end{lstlisting}
\end{tcolorbox}

\vspace{2mm}

{\ttfamily\small\color{codeyellow}input\_data/system/dimer\_electric\_dipoles.txt}
\vspace{1mm}
\begin{tcolorbox}[dampfterminal]
\begin{lstlisting}[style=dampfcode]
# x y z components of electric transition dipoles
1.0+0.0j   0.0+0.0j   0.0+0.0j
0.0+0.0j   1.0+0.0j   0.0+0.0j
\end{lstlisting}
\end{tcolorbox}

\vspace{2mm}

{\ttfamily\small\color{codeyellow}input\_data/system/dimer\_magnetic\_dipoles.txt}
\vspace{1mm}
\begin{tcolorbox}[dampfterminal]
\begin{lstlisting}[style=dampfcode]
# x y z components of magnetic transition dipoles
-1.0+0.0j   0.0+0.0j   1.0+0.0j
 0.0+0.0j  1.0+0.0j  -1.0+0.0j
\end{lstlisting}
\end{tcolorbox}
\end{tcolorbox}
\end{center}

The helper scripts \path{generate_system_hamiltonian.py} and \path{generate_dipole_moments.py} are provided in \path{helper_tools/} to generate these text files.

The pseudomode parameters are stored in \path{input_data/pseudomodes/}. A text file is required for each distinct local pseudomode environment. Thus, if all sites are coupled to identical copies of the same local pseudomode environment, a single file is sufficient. For site-dependent environments, one pseudomode-parameter file must be provided for each local environment.

Within these text files, each row specifies one pseudomode. The five columns contain, respectively, its frequency, damping rate, system--pseudomode coupling, thermal energy $k_{\mathrm B}T$ in the same energy units, and retained Fock-space dimension. The frequency and damping rate must be real and finite, with the frequency nonzero and the damping rate strictly positive. Both positive and negative frequencies are supported, but a negative-frequency pseudomode must have zero thermal energy. The coupling must be finite and can be complex, the thermal energy must be real, finite and non-negative, and the retained Fock-space dimension must be a positive integer. An example of a local pseudomode environment consisting of five pseudomodes is:

\begin{center}
\begin{tcolorbox}[dampffigure,title={Pseudomode input file}]
\centering
\begin{minipage}{0.78\linewidth}
\raggedright
{\ttfamily\small\color{codeyellow}\path{input_data/pseudomodes/Loc_env.txt}}
\vspace{1mm}
\begin{tcolorbox}[dampfterminal]
\begin{lstlisting}[style=dampfcode]
# frequency damping coupling thermal_energy fock_dim
400.0  10.0   20.0+0.0j   20.0   4
500.0  20.0   40.0+0.0j   20.0   4
600.0  15.0   70.0+0.0j   20.0   4
700.0  30.0   70.0+0.0j   20.0   4
800.0  10.0   50.0+0.0j   20.0   4
\end{lstlisting}
\end{tcolorbox}
\end{minipage}
\end{tcolorbox}
\end{center}

The helper script \path{generate_pseudomode_parameters.py} is provided in \path{helper_tools/} to generate these files.

\subsection{Configuration file}
\label{sec:configuration}

The file \path{configure_simulation.py} is the main user interface of DAMPyF, containing the parameters required to set up the simulation. A representative configuration is shown below:

\begin{center}
\begin{tcolorbox}[
    dampffigure,
    title={Representative configuration file},
    bottomrule=0pt,
    sharp corners=south
]
\begin{tcolorbox}[
    dampfterminal,
    bottomrule=0pt,
    sharp corners=south
]
{\ttfamily\small\color{codecream}
\begin{tabular}{@{}l@{}}

\textcolor{codegray}{\# options: "energy\_transfer", "linear\_spectra"}\\
simulation\_mode = "energy\_transfer"\\
\\
time = 100.0\\
dt = 1.0\\
dtdata = 1.0 * dt\\
backuptime\_seconds = 5 * 60 * 60\\
\\
\textcolor{codegray}{\# options: "first", "second"}\\
trotter\_order = "second"\\
\\
BD = 25\\
compression\_tol = 1.0e-8\\
\\
\textcolor{codegray}{\# options: "standard", "qr", "eig"}\\
compression\_svd\_method = "eig"\\
\\
system\_update\_coeff\_tol = 1.0e-8\\
\\
system\_hamiltonian\_file = "System\_hamiltonian.txt"\\
electric\_dipoles\_file = "Electric\_dipoles.txt"\\
magnetic\_dipoles\_file = "Magnetic\_dipoles.txt"\\

\end{tabular}
}
\end{tcolorbox}
\end{tcolorbox}
\end{center}

\clearpage

\begin{center}
\begin{tcolorbox}[
    dampffigure,
    toprule=0pt,
    sharp corners=north
]
\begin{tcolorbox}[
    dampfterminal,
    toprule=0pt,
    sharp corners=north
]
{\ttfamily\small\color{codecream}
\begin{tabular}{@{}l@{}}

same\_local\_environment = True\\
pseudomode\_parameter\_file = "Local\_env.txt"\\
pseudomode\_parameter\_files = ["Local\_env\_1.txt", "Local\_env\_2.txt"]\\
\\
\textcolor{codegray}{\# initial-state options:}\\
\textcolor{codegray}{\# "site", "eigenstate", "polarized\_pulse", "backup"}\\
initial\_state\_type = "site"\\
backup\_identifier = None\\
\\
initial\_site = 0\\
initial\_eigenstate = 1\\
light\_polarization = np.array([1.0, 1.0, 1.0], dtype=float)\\
\\
output\_identifier = "example\_simulation"\\
\\
\textcolor{codegray}{\# options: "local", "slurm"}\\
execution\_mode = "local"\\
\\
ray\_max\_parallel\_system\_updates = None\\

\end{tabular}
}
\end{tcolorbox}
\end{tcolorbox}
\end{center}

The main configuration variables are:
\begin{itemize}

\item \path{simulation_mode}: selects the workflow to run. The available options are \path{"energy_transfer"}, in which an initial system density matrix is propagated in time, and \path{"linear_spectra"}, tailored to linear-spectra simulations of molecular systems, in which site-resolved optical coherences are propagated; see Appendix~\ref{app_spectra} for further details.

\item \path{time}: total propagation time, expressed in the time unit conjugate to the energy unit used for the system and pseudomode parameters.

\item \path{dt}: propagation time step, expressed in the same units as \path{time}.

\item \path{dtdata}: time interval between successive stored data points. It must be an integer multiple of \path{dt}, and \path{time} must in turn be an integer multiple of \path{dtdata}.

\item \path{backuptime_seconds}: wall-clock interval, expressed in seconds, between successive checkpoint operations. When this interval is reached, DAMPyF stores the output data collected so far and saves the evolved MPS components in a checkpoint folder.

\item \path{trotter_order}: selects the Trotter decomposition used for the propagation. The available options are \path{"first"} and \path{"second"}, corresponding, respectively, to the first-order and symmetric second-order decompositions discussed in Sec.~\ref{sec:time_evolution}.

\item \path{BD}: maximum MPS bond dimension retained during compression; see Appendix~\ref{app:compression} for further details.

\item \path{compression_tol}: relative single-tensor truncation tolerance used during MPS compression. Together with \path{BD}, it controls the effective bond dimensions retained at each time step; see Appendix~\ref{app:compression} for further details.

\item \path{compression_svd_method}: selects the numerical backend used for MPS compression. The available options are \path{"standard"}, \path{"qr"}, and \path{"eig"}; see Appendix~\ref{app:compression} for further details.

\item \path{system_update_coeff_tol}: threshold used to neglect matrix elements of the system propagator \(U_{\mathrm{s}}(dt)\) whose absolute values fall below it during the computationally costly system update implementing Eq.~\eqref{eq:system_mixing}.

\item \path{system_hamiltonian_file}: name of the system-Hamiltonian file stored in \path{input_data/system/}.

\item \path{electric_dipoles_file}: name of the optional electric-transition-dipole file stored in \path{input_data/system/}.

\item \path{magnetic_dipoles_file}: name of the optional magnetic-transition-dipole file stored in \path{input_data/system/}.

\item \path{same_local_environment}: specifies whether all system sites are coupled to identical copies of the same local pseudomode environment.

\item \path{pseudomode_parameter_file}: name of the pseudomode-parameter file used when \path{same_local_environment} is set to \path{True}. The file is stored in \path{input_data/pseudomodes/} and defines the local environment replicated independently at every site.

\item \path{pseudomode_parameter_files}: list containing the names of the pseudomode-parameter files used when \path{same_local_environment} is set to \path{False}. The files are stored in \path{input_data/pseudomodes/} and define the local environments associated with the individual sites. Use \path{None} for sites without local environments.

\item \path{initial_state_type}: selects the initial state. The available options are \path{"site"}, \path{"eigenstate"}, \path{"polarized_pulse"}, and \path{"backup"}.

For the first three options, the initial system state is
\begin{equation*}
    \rho_{\mathrm{s}}(0)
    =
    |\psi\rangle\!\langle\psi|.
\end{equation*}

For \path{"site"}, the initial state is
\begin{equation*}
    |\psi\rangle=|n_0\rangle,
\end{equation*}
where \(|n_0\rangle\) is the initially populated site selected through \path{initial_site}.

For \path{"eigenstate"}, the initial state is
\begin{equation*}
    |\psi\rangle
    =
    |\varepsilon_\alpha\rangle,
    \qquad
    H_{\mathrm{s}}|\varepsilon_\alpha\rangle
    =
    E_\alpha|\varepsilon_\alpha\rangle,
\end{equation*}
where the eigenstates are ordered by increasing energy and the desired eigenstate is selected through \path{initial_eigenstate}.

For \path{"polarized_pulse"}, the initial state is constructed from the electric transition dipoles provided through \path{electric_dipoles_file} and the light-polarization vector \(\boldsymbol{e}\) specified through \path{light_polarization}, according to
\begin{equation*}
    |\psi\rangle
    =
    \frac{1}{\sqrt{\mathcal{N}}}
    \sum_{n=0}^{N-1}
    \left(
    \boldsymbol{e}\cdot\boldsymbol{\mu}_n
    \right)
    |n\rangle,
\end{equation*}
where \(\mathcal{N}\) is the normalization factor

\begin{equation*}
\mathcal{N}
=
\sum_{n=0}^{N-1}
\left|
\boldsymbol{e}\cdot\boldsymbol{\mu}_n
\right|^2.
\end{equation*}

For \path{"backup"}, the MPS components are loaded from the checkpoint folder selected through \path{backup_identifier}. The loaded MPSs are used as a new initial state.

\item \path{backup_identifier}: exact name of the checkpoint folder, located in \path{output_data/}, loaded when \path{initial_state_type} is set to \path{"backup"}. The checkpoint must be compatible with the selected simulation mode, system size, and pseudomode Fock-space dimensions.

\item \path{output_identifier}: nonempty identifier string appended to the names of the output files generated by the simulation.

\item \path{execution_mode}: selects the execution environment. For \path{"local"}, DAMPyF is launched directly through the \path{run_dampf.py} file. The \path{"slurm"} option is used on a Slurm-managed platform, where DAMPyF can be launched through the \path{slurm_code_launcher.sh} script after the user specifies the requested resources. The memory required by DAMPyF simulations can be roughly estimated using the helper script \path{estimate_required_memory.py} in \path{helper_tools/}.

\item \path{ray_max_parallel_system_updates}: advanced parameter specifying an upper bound on the number of simultaneous parallel tasks used during the system-update step, which is generally the most memory-intensive operation in DAMPyF. It can be set to \path{None} or to a positive integer. When set to \path{None}, DAMPyF determines the initial value automatically. In either case, the actual number of simultaneous tasks is automatically reduced if the estimated memory requirement exceeds the available memory.
\end{itemize}

The variables not used for a given simulation can be set to \path{None}.

\subsection{Output data}
\label{sec:outputs}

Simulation outputs are written to \path{output_data/}. The output filenames contain the value of \path{output_identifier}, allowing different runs to be stored in the same folder.

For \path{output_identifier = "test_simulation"}, the core output files are:

\begin{center}
\begin{tcolorbox}[dampffigure,title={Output files}]
\centering

\begin{minipage}{0.78\linewidth}
\raggedright
{\ttfamily\small\color{codeteal}energy-transfer mode}
\vspace{1mm}
\begin{tcolorbox}[dampfterminal]
\begin{lstlisting}[style=dampfcode]
output_data/
|-- rho_system_test_simulation.npz
`-- simulation_data_test_simulation.txt
\end{lstlisting}
\end{tcolorbox}
\end{minipage}

\vspace{2mm}

\begin{minipage}{0.78\linewidth}
\raggedright
{\ttfamily\small\color{codeteal}linear-spectra mode}
\vspace{1mm}
\begin{tcolorbox}[dampfterminal]
\begin{lstlisting}[style=dampfcode]
output_data/
|-- optical_coherence_test_simulation.npz
`-- simulation_data_test_simulation.txt
\end{lstlisting}
\end{tcolorbox}
\end{minipage}

\end{tcolorbox}
\end{center}

\subsubsection{Core simulation output}

For energy-transfer simulations, the file
\path{rho_system_<output_identifier>.npz}
contains the arrays \path{times} and \path{rho_system}. Given
\begin{equation*}
    N_t
    =
    \operatorname{round}
    \left(
    \frac{\mathtt{time}}{\mathtt{dtdata}}
    \right)
    +1
\end{equation*}
stored time points, \path{times} has shape \((N_t,)\), while \path{rho_system} has shape \((N_t,N,N)\). The first axis of \path{rho_system} labels the sampled times, while the remaining axes contain the corresponding reduced system density matrices.

For linear-spectra simulations, the file
\path{optical_coherence_<output_identifier>.npz}
contains the arrays \path{times} and \path{optical_coherence}. The latter has shape \((N_t,N,N)\). At each sampled time, the element \((m,n)\) contains the evolved optical coherence on site \(n\) obtained by taking the optical coherence on site \(m\) as the initial state; see Appendix~\ref{app_spectra} for further details.

The file \path{simulation_data_<output_identifier>.txt} is intended to record the information identifying the simulation. It contains the main propagation settings, initial-state information, system Hamiltonian, optional electric and magnetic transition dipoles, and pseudomode parameters. Additionally, it records the total wall-clock time required to complete the simulation, the maximum bond dimension reached during the evolution, and the maximum bond dimension at the final time.

\subsubsection{Plotting and spectral-analysis helpers}

The helper script \path{analysis_tools/plot_system_dynamics.py} loads the time axis and reduced system density matrices from an energy-transfer output file and can be used to plot populations and coherences in either the site or eigenstate basis.

The scripts \path{analysis_tools/compute_absorption_spectrum.py} and
\path{analysis_tools/compute_circular_dichroism_spectrum.py} load the time axis and optical coherences from a linear-spectra output file and compute the corresponding spectra. The processed data are written to \path{output_data/}.

\subsubsection{Checkpoint data}

During long calculations, DAMPyF may create checkpoint folders of the form
\path{checkpoint_<output_identifier>_step_<step_index>/},
according to the value of \path{backuptime_seconds} and the time step reached at the moment of the backup. When a checkpoint is created, the output data collected up to that point are saved in the corresponding \path{.npz} file, while the current MPS components are stored as \path{.hdf5} files inside the checkpoint folder.

For energy-transfer simulations, the checkpoint files have the form
\path{state_<component_index>.hdf5}. For linear-spectra simulations, they have the form
\path{state_<initial_index>_<current_index>.hdf5}. These checkpoint folders can subsequently be selected through \path{backup_identifier} and used as the initial state of a new propagation.

\section{Installation and execution}
\label{sec:installation}

\subsection{Requirements}

DAMPyF is a pure-Python project built on standard scientific Python packages. It requires
Python 3, NumPy, SciPy, h5py, psutil, and Ray for parallelization. Matplotlib and a LaTeX
installation are additionally required by the plotting tools. The software versions used to
test DAMPyF on Windows and Linux are listed in Appendix~\ref{sec:tested-software-versions}.

\subsection{Execution}

Before launching a calculation, the user should edit \path{configure_simulation.py} and provide the relevant input parameter files. DAMPyF can be executed either locally or within a Slurm allocation, as in:

\begin{center}
\begin{tcolorbox}[dampffigure,title={Running DAMPyF}]
\begin{minipage}{0.47\linewidth}
{\ttfamily\small\color{codeteal}Local execution}
\begin{tcolorbox}[dampfterminal]
\begin{lstlisting}[style=dampfcode]
python run_dampf.py
\end{lstlisting}
\end{tcolorbox}
\end{minipage}
\hfill
\begin{minipage}{0.47\linewidth}
{\ttfamily\small\color{codeteal}Slurm execution}
\begin{tcolorbox}[dampfterminal]
\begin{lstlisting}[style=dampfcode]
sbatch slurm_code_launcher.sh
\end{lstlisting}
\end{tcolorbox}
\end{minipage}
\end{tcolorbox}
\end{center}

For local calculations, \path{execution_mode} must be set to \path{"local"} in \path{configure_simulation.py}. For Slurm calculations, it must instead be set to \path{"slurm"}, and the requested CPUs and memory must be specified in \path{slurm_code_launcher.sh}. The supplied Slurm launcher template assumes that a Conda environment named \path{DAMPyF_environment}, containing the required packages, has been created and can be activated. DAMPyF reads the allocated resources from the Slurm environment and uses Ray within the allocated node.

Before long calculations, the helper script \path{helper_tools/estimate_required_memory.py} can be used to obtain a rough estimate of the memory required by the current configuration.

\section{Convergence parameters}
\label{sec:convergence}

The main convergence parameters in a DAMPyF calculation are:

\begin{itemize}
\item propagation time step \path{dt};
\item maximum bond dimension \path{BD};
\item pseudomode Fock-space dimensions;
\item compression tolerance \path{compression_tol};
\item system-update coefficient tolerance \path{system_update_coeff_tol}.
\end{itemize}

We recommend setting \path{compression_tol} and \path{system_update_coeff_tol} to sufficiently small values to ensure convergence over the simulated timescale. The convergence analysis can then focus primarily on the propagation time step, maximum bond dimension, and pseudomode Fock-space dimensions.

The helper script \path{helper_tools/estimate_pseudomode_fock_dimensions.py} can be used to obtain a rough estimate of the pseudomode Fock-space dimensions required for the simulation.

\section{Minimal examples}
\label{sec:examples}

The example in \path{Examples/Dimer_energy_transfer/} considers a dimeric system with \(N=2\). The system Hamiltonian is provided in \path{dimer_system_hamiltonian.txt}, while each site is coupled to an identical copy of the five-pseudomode environment specified in \path{Loc_env.txt}. These input files correspond to the examples introduced in Sec.~\ref{sec:input_data}.

The Hamiltonian and pseudomode parameters are expressed in \(\mathrm{cm}^{-1}\). Times specified in femtoseconds are converted using \path{FEMTOSECOND_TO_SPECTROSCOPIC_TIME}, defined in \path{dampf_modules/unit_conventions.py}.

\subsection{Dimer energy-transfer dynamics}
\label{sec:example_1}

We consider energy-transfer dynamics starting from the higher-energy system eigenstate. The example \path{configure_simulation.py} file contains the following principal settings:

\begin{center}
\begin{tcolorbox}[dampffigure,title={Dimer energy-transfer configuration file}]
\begin{tcolorbox}[dampfterminal]
{\ttfamily\small\color{codecream}
\begin{tabular}{@{}l@{}}

simulation\_mode = "energy\_transfer"\\
\\
time = 200.0*units.FEMTOSECOND\_TO\_SPECTROSCOPIC\_TIME\\
dt = 1.0*units.FEMTOSECOND\_TO\_SPECTROSCOPIC\_TIME\\
dtdata = 1.0 * dt\\
backuptime\_seconds = 5 * 60 * 60\\
trotter\_order = "second"\\
\\
BD = 5\\
compression\_tol = 1.0e-8\\
compression\_svd\_method = "eig"\\
system\_update\_coeff\_tol = 1.0e-8\\
\\
system\_hamiltonian\_file = "dimer\_system\_hamiltonian.txt"\\
electric\_dipoles\_file = None\\
magnetic\_dipoles\_file = None\\
\\
same\_local\_environment = True\\
pseudomode\_parameter\_file = "Loc\_env.txt"\\
pseudomode\_parameter\_files = []\\
\\
initial\_state\_type = "eigenstate"\\
backup\_identifier = None\\
initial\_site = None\\
initial\_eigenstate = 1\\
light\_polarization = None\\
\\
output\_identifier = "dimer\_test\_simulation"\\
\\
execution\_mode = "local"\\
ray\_max\_parallel\_system\_updates = None\\

\end{tabular}
}
\end{tcolorbox}
\end{tcolorbox}
\end{center}

The calculation is launched by running the Python script \path{run_dampf.py}. It generates the files \path{rho_system_dimer_test_simulation.npz} and \path{simulation_data_dimer_test_simulation.txt} in \path{output_data/}.

The helper script \path{analysis_tools/plot_system_dynamics.py} can be used to plot the site populations and inter-site coherences. It can also compare two datasets, which is useful for convergence checks. For example, the calculation above can be compared with a second simulation obtained by changing the following parameters:

\begin{center}
\begin{tcolorbox}[dampffigure,title={Dimer energy-transfer configuration file}]
\begin{tcolorbox}[dampfterminal]
{\ttfamily\small\color{codecream}
\begin{tabular}{@{}l@{}}
BD = 15\\
dt = 0.5*units.FEMTOSECOND\_TO\_SPECTROSCOPIC\_TIME\\
output\_identifier = "dimer\_test\_simulation\_2"
\end{tabular}
}
\end{tcolorbox}
\end{tcolorbox}
\end{center}

\begin{figure}[H]
    \centering
    \includegraphics[width=\linewidth]{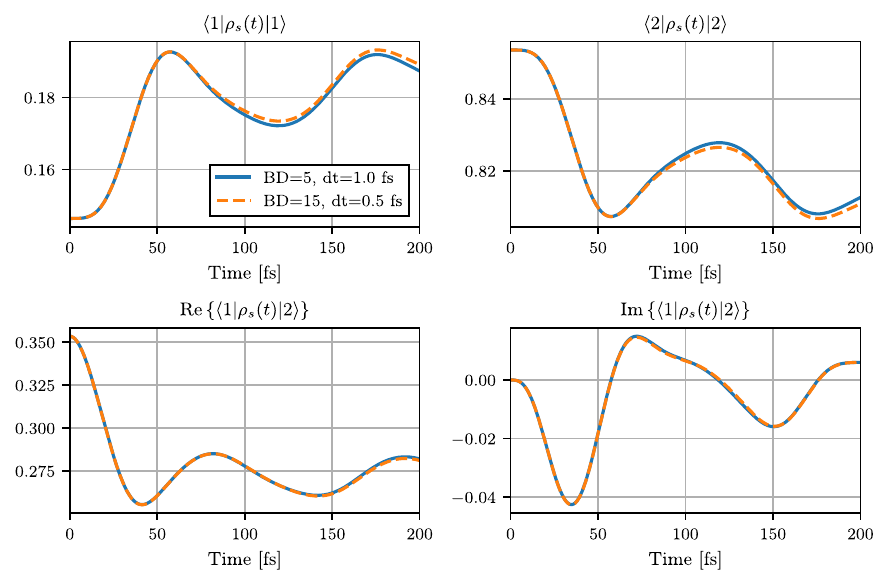}
    \caption{Comparison of the dimer energy-transfer dynamics obtained using the two numerical parameter sets described in the text.}
    \label{fig:dimer_energy_transfer_example}
\end{figure}

An example comparison generated via \path{analysis_tools/plot_system_dynamics.py} is shown in Fig.~\ref{fig:dimer_energy_transfer_example}, proving that the set of parameters used for the first dataset does not suffice to reach convergence.

\subsection{Dimer linear spectra}

We now consider the linear spectra of the dimeric unit. The example \path{configure_simulation.py} file is the same as that used in Sec.~\ref{sec:example_1}, with only the following differences: 

\begin{center}
\begin{tcolorbox}[dampffigure,title={Dimer linear-spectra configuration file}]
\begin{tcolorbox}[dampfterminal]
{\ttfamily\small\color{codecream}
\begin{tabular}{@{}l@{}}

simulation\_mode = "linear\_spectra"\\
\\
initial\_state\_type = None\\
initial\_eigenstate = None\\

\end{tabular}
}
\end{tcolorbox}
\end{tcolorbox}
\end{center}

The calculation is launched by running the command \path{python run_dampf.py}. It generates the files \path{optical_coherence_dimer_test_simulation.npz} and \path{simulation_data_dimer_test_simulation.txt} in \path{output_data/}.

The analysis scripts \path{analysis_tools/compute_absorption_spectrum.py} and \path{analysis_tools/compute_circular_dichroism_spectrum.py} can be used to plot the absorption and circular dichroism spectra, respectively. These scripts require the specification of \path{electric_dipoles_file} and \path{magnetic_dipoles_file}, here taken as

\begin{center}
\begin{tcolorbox}[dampffigure,title={Compute absorption/circular dichroism spectra files}]
\begin{tcolorbox}[dampfterminal]
{\ttfamily\small\color{codecream}
\begin{tabular}{@{}l@{}}

electric\_dipoles\_file = "dimer\_electric\_dipoles.txt"  \\
magnetic\_dipoles\_file = "dimer\_magnetic\_dipoles.txt" \\

\end{tabular}
}
\end{tcolorbox}
\end{tcolorbox}
\end{center}

whose values are given in Sec.~\ref{sec:input_data}.

\begin{figure}[H]
    \centering
    \includegraphics[width=\linewidth]{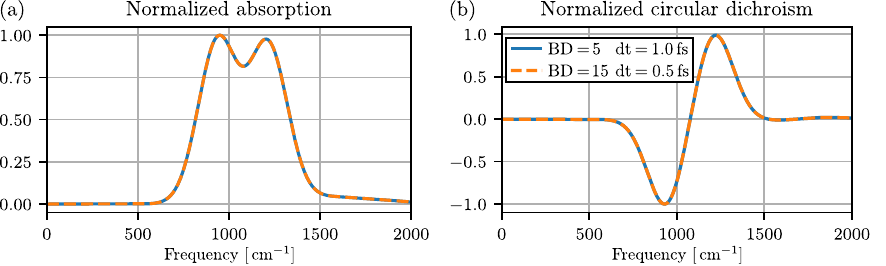}
    \caption{Comparison of the dimer absorption and circular dichroism spectra obtained with the two sets of numerical parameters described in the text. A Gaussian filter with $\sigma=50\,\mathrm{fs}$ was applied to the computed optical coherences to ensure their decay within the simulated time interval of $200\,\mathrm{fs}$.}
    \label{fig:dimer_linear_spectra}
\end{figure}

The analysis scripts can also compare two datasets, which is useful for convergence checks. We consider the same two sets of convergence parameters used for the dimer energy-transfer dynamics in Sec.~\ref{sec:example_1}. The resulting comparison is shown in Fig.~\ref{fig:dimer_linear_spectra}, which indicates that, for the linear spectra, even the first dataset leads to converged results. For this comparison, the plotting scripts in \path{analysis_tools/} were configured as follows:

\begin{center} 
\begin{tcolorbox}[dampffigure,title={Compute absorption/circular dichroism spectra files}]
\begin{tcolorbox}[dampfterminal] 
{\ttfamily\small\color{codecream} 
\begin{tabular}{@{}l@{}} 
frequency\_shift = 0.0\\
frequency\_min = 0.0\\
frequency\_max = 2000.0\\ 
frequency\_points = 6000\\ 
\\
interpolation\_factor = 5\\ 
apply\_gaussian\_filter = True\\ 
gaussian\_sigma = 50.0*units.FEMTOSECOND\_TO\_SPECTROSCOPIC\_TIME\\ 
normalize\_spectrum = True\\ 
\end{tabular} 
} 
\end{tcolorbox} 
\end{tcolorbox} 
\end{center}

The first four parameters specify an optional shift of the resulting frequency axis, the frequency range over which the Fourier transform is evaluated before this shift is applied, and the number of frequency points generated within this range.

The remaining parameters control preprocessing before the Fourier transform and normalization of the spectra. The parameter \path{interpolation_factor} increases the number of time points in the correlation function through interpolation before the Fourier transformation. When \path{apply_gaussian_filter} is set to \path{True}, a Gaussian time-domain filter with width \path{gaussian_sigma} is applied. Here, the width is chosen such that the filtered correlation function decays within the simulated time interval, thereby suppressing ringing artifacts caused by finite-time truncation. Finally, \path{normalize_spectrum} determines whether the resulting spectrum is normalized.

\section{Conclusion}
\label{sec:conclusion}

DAMPyF provides an open-source Python implementation of the DAMPF method for non-perturbative simulations of multi-site systems coupled to local pseudomode environments. Its self-contained, pure-Python codebase implements the core DAMPF functionality together with an extension for calculating linear spectra.

Its open-source, modular structure also makes DAMPyF suitable for several possible model extensions, including support for non-local pseudomode environments~\cite{SanchezBarquillaNanophotonics2022}, interactions between pseudomodes~\cite{MascherpaPRA2020,MedinaPRL2021,parkPRB2024,huangPRL2026}, non-Hermitian pseudomode models~\cite{MenczelPRR2024}, nonlinear system--environment couplings~\cite{ZhangPRB2025}, and additional dissipative processes representing coupling to large external baths through Lindblad terms, as required, for example, to model irreversible energy transfer via sinks~\cite{CarusoJCP2009,ChinNJP2010} or coupling to hot and cold reservoirs~\cite{StrasbergNJP2016}. 

DAMPyF is therefore an appealing tool for simulations of systems coupled to structured environments, ranging from electronic dynamics in molecular systems~\cite{SomozaCommunPhys2023, LorenzoniPRL2024,LorenzoniSciAdv2025} and the dynamics of quantum emitters coupled to structured photonic environments~\cite{PriorPRA2013, MedinaPRL2021} to the benchmarking of quantum simulations of open-system models implemented on trapped-ion platforms, where the motional phonons realize the environmental degrees of freedom~\cite{Lemmer_2018,SoSciAdv2024,SunNC2025}, superconducting circuits, where engineered microwave modes realize structured environments~\cite{MostameIOP2012, MagazzuNC2018, LeppakangasPRA2018}, and Rydberg-atom platforms, where atomic interactions couple the internal electronic states to atomic vibrational modes~\cite{MagoniPRL2023, EuchnerPRR2025, ZhangPRA2025}.

\section*{Acknowledgments}

We thank Namgee Cho for carefully proofreading the manuscript and testing the DAMPyF code.

\paragraph{Funding information}
This work was supported by the ERC Synergy Grant HyperQ (Grant No.~856432), the BMBF project PhoQuant (Grant No.~13N16110), and the State of Baden-Württemberg through bwHPC and the German Research Foundation (DFG) through Grant No.~INST 40/575-1 FUGG (JUSTUS 2 cluster).

\begin{appendix}
\numberwithin{equation}{section}

\section{Tested software versions}
\label{sec:tested-software-versions}
DAMPyF was tested on Windows and Linux using the software versions listed in
Table~\ref{tab:tested-software-versions}.

\begin{table}[tbh]
\centering
\begin{tabular}{lcc}
\hline
Package & Windows & Linux \\
\hline
Python     & 3.12.13 & 3.14.6 \\
NumPy      & 2.5.1   & 2.5.1  \\
SciPy      & 1.18.0  & 1.18.0 \\
Ray        & 2.56.1  & 2.56.1 \\
h5py       & 3.16.0  & 3.16.0 \\
psutil     & 7.2.2   & 7.2.2  \\
Matplotlib & 3.11.1  & 3.11.1 \\
\hline
\end{tabular}
\caption{Software versions used to test DAMPyF on Windows and Linux.}
\label{tab:tested-software-versions}
\end{table}

\section{Characterization of pseudomode parameters}
\label{app:pseudomode_parameters}

DAMPyF takes the pseudomode parameters as input.

Pseudomode environments of the form introduced in the main text can be used as effective representations of continuous bosonic environments~\cite{TamascelliPRL2018}.

Consider a target model with Hamiltonian
\begin{equation}
\label{eq:app_target_hamiltonian}
H
=
H_{\mathrm{s}}
+
H_{\tilde{\mathrm{e}}}
+
H_{\mathrm{s-\tilde{e}}},
\end{equation}
where \(H_{\mathrm{s}}\) is the system Hamiltonian defined in Eq.~\eqref{eq:hamiltonian_system}, while the environment and system--environment coupling Hamiltonians are
\begin{equation}
\label{eq:app_target_environment}
\begin{aligned}
H_{\tilde{\mathrm{e}}}
&=
\sum_{n=0}^{N-1}
\int d\omega\,
\omega\,
a_n^{\dagger}(\omega)a_n(\omega),
\\
H_{\mathrm{s-\tilde{e}}}
&=
\sum_{n=0}^{N-1}
|n\rangle\!\langle n|
\otimes
\int d\omega\,
\left[g_n(\omega)
a_n^{\dagger}(\omega)+g_n(\omega)^*a_n(\omega)
\right].
\end{aligned}
\end{equation}
Here, \(a_n^{\dagger}(\omega)\) and \(a_n(\omega)\) are the creation and annihilation operators of the bosonic mode of frequency \(\omega\) belonging to the environment coupled to site \(n\), while \(g_n(\omega)\) denotes its coupling strength. The dynamics of the total system is governed by
\begin{equation}
\label{eq:app_target_eom}
\frac{d}{dt}\rho_{\mathrm{s\tilde{e}}}(t)
=
-\mathrm{i}
\left[
H,\rho_{\mathrm{s\tilde{e}}}(t)
\right].
\end{equation}

The continuous environments can be replaced by pseudomode environments described by the Hamiltonian and equation of motion introduced in Eqs.~\eqref{eq:hamiltonian_pseudomodes}--\eqref{eq:dissipator}. If the target and pseudomode environments induce identical bath correlation functions, they generate the same reduced system dynamics and the replacement is therefore numerically exact~\cite{TamascelliPRL2018}. Under the corresponding Gaussianity assumptions, this equivalence also extends to multi-time expectation values~\cite{SmirneOSID2022}.

For each local environment, the pseudomode bath correlation function is
\begin{equation}
\label{eq:app_pseudomode_bcf}
C_{\mathrm{ps},n}(t)
=
\sum_{q=0}^{Q_n-1}
|g_{nq}|^{2}
e^{-\gamma_{nq}t/2}
\left[
\left(1+\bar n_{nq}\right)e^{-\mathrm{i}\omega_{nq}t}
+
\bar n_{nq}e^{\mathrm{i}\omega_{nq}t}
\right],
\end{equation}
where \(\omega_{nq}\), \(\gamma_{nq}\), \(g_{nq}\), and \(\bar n_{nq}\) are, respectively, the frequency, damping rate, coupling strength, and thermal occupation of pseudomode \(q\) coupled to site \(n\).

The pseudomode parameters must be chosen such that \(C_{\mathrm{ps},n}(t)\) reproduces the target bath correlation function
\begin{equation}
\label{eq:app_target_bcf}
C_n(t)
=
\int_{-\infty}^{\infty}d\omega\,
e^{-\mathrm{i}\omega t}C_n(\omega),
\end{equation}

with 
\begin{equation*}
C_n(\omega)=[1+n(\omega)]J_n(\omega)\Theta(\omega)+n(|\omega|)J_n(|\omega|)\Theta(-\omega),
\end{equation*}

where \(n(\omega)\) is the Bose occupation factor, $\Theta(\omega)$ the Heaviside step function and the spectral density associated with the \(n\)-th environment is
\begin{equation}
\label{eq:app_target_spectral_density}
J_n(\omega)
=
\int d\omega'\,
|g_n(\omega')|^{2}\,
\delta(\omega-\omega').
\end{equation}

In practice, the matching may be performed over the finite time interval relevant to the target dynamics, allowing systematic coarse graining of the environment~\cite{LorenzoniPRL2024}. Bounds on the resulting dynamical error can be expressed in terms of the bath-correlation-function mismatch~\cite{MascherpaPRL2017}. The construction and fitting of the pseudomode parameters required for this representation are external to DAMPyF.

\section{Linear absorption and circular dichroism spectra}
\label{app_spectra}

The linear-spectra workflow in DAMPyF considers molecular systems and propagates optical coherences between the electronic ground state and the single-excitation manifold. Let $\left|g\right\rangle$ be the electronic ground state and $\left|e_n\right\rangle$ the state with a local excitation on site $n$. The electric transition-dipole operator is written as
\begin{equation}
\label{eq:app_electric_dipole_operator}
\boldsymbol{\mu}
=
\sum_{n=0}^{N-1}
\left[
\boldsymbol{\mu}_n
\left|e_n\right\rangle\!\left\langle g\right|
+
\boldsymbol{\mu}_n^{*}
\left|g\right\rangle\!\left\langle e_n\right|
\right],
\end{equation}
where $\boldsymbol{\mu}_n$ is the electric transition dipole of site $n$. If circular dichroism is considered, one also introduces the magnetic transition-dipole operator
\begin{equation}
\label{eq:app_magnetic_dipole_operator}
\boldsymbol{m}
=
\sum_{n=0}^{N-1}
\left[
\boldsymbol{m}_n
\left|e_n\right\rangle\!\left\langle g\right|
+
\boldsymbol{m}_n^{*}
\left|g\right\rangle\!\left\langle e_n\right|
\right].
\end{equation}
The helper script for generating dipoles also permits the use of the point-dipole convention
\begin{equation}
\label{eq:app_magnetic_helper_convention}
\boldsymbol{m}_n=\boldsymbol{r}_n\times\boldsymbol{\mu}_n,
\end{equation}
where $\boldsymbol{r}_n$ is the position of site $n$. Absolute prefactors depend on the unit convention used for the magnetic dipoles; in DAMPyF post-processing, the spectra are usually normalized, so only the relative lineshape is relevant.

We represent the optical-coherence response by an array $\eta_{ij}(t)$, where the first index labels the initially created optical coherence and the second index labels the final excited-state component after propagation. In terms of the full system-plus-pseudomode evolution,
\begin{equation}
\label{eq:app_optical_coherence_definition}
\eta_{ij}(t)=
\operatorname{Tr}_{\mathrm e}\!
\left[
\left\langle e_j\right|
 e^{\mathcal{L}t}
\left(
\left|e_i\right\rangle\!\left\langle g\right|\otimes\rho_{\mathrm{e}}^{\mathrm{th}}
\right)
\left|g\right\rangle
\right],
\end{equation}
where the trace is over the pseudomode degrees of freedom and $\rho_{\mathrm{e}}^{\mathrm{th}}$ is the product of thermal pseudomode states.

Most importantly, because the electronic ground state is decoupled from the excited-state manifold, the evolution of an optical coherence remains confined to the $N$-dimensional subspace spanned by the optical coherences ($\{|e_i\rangle\langle g|\}_{i=0}^{N-1}$). Taking the ground-state energy as the energy reference, the dynamics within this subspace is fully determined by the excited-state Hamiltonian introduced in the main text; see Eq.~\eqref{eq:hamiltonian_system}. Thus, no additional ground-state degrees of freedom need to be included.

For an isotropic ensemble, the rotationally averaged electric-dipole correlation used for absorption is
\begin{equation}
\label{eq:app_dipole_correlation_code}
C_{\mu\mu}(t)=\frac{1}{3}
\sum_{i,j}
\boldsymbol{\mu}_j^{*}\cdot\boldsymbol{\mu}_i\,
\eta_{ij}(t),
\end{equation}
which matches the convention implemented in the absorption post-processing script.

The absorption spectrum is then computed from the real part of a one-sided Fourier transform~\cite{Mukamel1995},
\begin{equation}
\label{eq:app_absorption_spectrum}
A(\omega)\propto
\operatorname{Re}
\int_0^{\infty}dt\,
 e^{\mathrm{i}\omega t}C_{\mu\mu}(t).
\end{equation}
In practical calculations, the signal obtained over the finite simulated time window can be multiplied by a filter function $F(t)$ to ensure sufficient decay within the simulated timescale and thereby suppress ringing artifacts. For instance, using a Gaussian filter gives
\begin{equation}
\label{eq:app_filtered_absorption_spectrum}
A(\omega)\propto
\operatorname{Re}
\int_0^{t_{\max}}dt\,
 e^{\mathrm{i}\omega t}F(t)C_{\mu\mu}(t),
\qquad
F(t)=\exp\left(-\frac{t^2}{2\sigma^2}\right).
\end{equation}

A circular-dichroism spectrum is obtained from the mixed electric--magnetic response~\cite{DinhRengerJCP2015}. In the same optical-coherence convention, the correlation function implemented in \path{compute_circular_dichroism_spectrum.py} is
\begin{equation}
\label{eq:app_cd_correlation}
C_{m\mu}(t)=\frac{1}{3}
\sum_{i,j}
\boldsymbol{m}_j^{*}\cdot\boldsymbol{\mu}_i\,
\eta_{ij}(t).
\end{equation}
This is the same contraction used for absorption, but with the first electric dipole replaced by the corresponding magnetic transition dipole. The circular-dichroism lineshape is obtained from the real part of the one-sided Fourier transform,
\begin{equation}
\label{eq:app_cd_spectrum}
CD(\omega)\propto
\operatorname{Re}
\int_0^{\infty}dt\,
 e^{\mathrm{i}\omega t}C_{m\mu}(t).
\end{equation}
Equivalently, with the same finite-time filtering used for absorption,
\begin{equation}
\label{eq:app_filtered_cd_spectrum}
CD(\omega)\propto
\operatorname{Re}
\int_0^{t_{\max}}dt\,
 e^{\mathrm{i}\omega t}F(t)C_{m\mu}(t).
\end{equation}

\section{Compression algorithm}
\label{app:compression}

The application of the system propagator involves sums of MPSs and therefore increases their bond dimensions. It must therefore be accompanied by a compression step, which in DAMPyF is performed using a singular-value decomposition (SVD) compression algorithm. This algorithm performs a sweep of SVDs along the MPS, compressing each tensor to the desired bond dimension. The compression requires an initial canonicalization of the MPS, followed by the actual compression sweep. In the following, we assume that the reader is familiar with these basic concepts regarding MPSs and otherwise refer to the relevant literature~\cite{SchollwockAnnPhys2011,OrusAnnPhys2014}.

To illustrate the procedure, we consider the case in which the bond dimension of the MPS has already saturated the maximum allowed value $\chi=$\path{BD}. The sum produced by the system propagator then results, after each summation step, in an MPS with bond dimension $2\chi$, which must be compressed to a bond dimension no larger than $\chi$.

Consider a compression sweep from left to right. During this tensor-by-tensor compression, the algorithm considers an uncompressed bulk tensor $A^{(m,n)}_{\ell,i_\ell}$ of shape $\chi\times d_\ell^2\times 2\chi$. This tensor is reshaped into a rectangular matrix $M^{(m,n)}_\ell$ of shape $\chi d_\ell^2\times 2\chi$. Its SVD gives
\begin{equation}
\label{eq:app_svd}
M^{(m,n)}_\ell
=
U^{(m,n)}_\ell
S^{(m,n)}_\ell
V^{(m,n)\dagger}_\ell,
\end{equation}
where $U^{(m,n)}_\ell$, $S^{(m,n)}_\ell$, and $V^{(m,n)\dagger}_\ell$ have shapes $\chi d_\ell^2\times\alpha$, $\alpha\times\alpha$, and $\alpha\times 2\chi$, respectively, with $\alpha=\min\{\chi d_\ell^2,2\chi\}$. Once the SVD is performed, the compression discards the sectors of these three matrices associated with the least significant singular values.

In particular, let $s_i$ denote the singular values in decreasing order, let $\varepsilon_{\mathrm c}$ denote the value of \path{compression_tol}, and let $r_{\mathrm{tol}}$ be the smallest rank satisfying $\sum_{i=1}^{r_{\mathrm{tol}}}s_i/\sum_i s_i\geq 1-\varepsilon_{\mathrm c}$. The retained rank is $r=\min\{\chi_{\max},r_{\mathrm{tol}}\}$, where \(\chi_{\max}=\mathtt{BD}\). Thus, for the illustrative tensor above, the truncated matrices $U^{(m,n)}_\ell$, $S^{(m,n)}_\ell$, and $V^{(m,n)\dagger}_\ell$ have shapes $\chi d_\ell^2\times r$, $r\times r$, and $r\times 2\chi$, respectively. By reshaping $U^{(m,n)}_\ell$ into the new local tensor and absorbing the product $S^{(m,n)}_\ell V^{(m,n)\dagger}_\ell$ into the next tensor, the corresponding bond is reduced to the selected rank, and the compression sweep can proceed to the next site.

DAMPyF provides three methods for computing the decomposition of the non-square matrix $M^{(m,n)}_\ell$, selected through \path{compression_svd_method}. All three methods produce the factors entering Eq.~\eqref{eq:app_svd}, but use different intermediate decompositions.

For \path{standard}, the singular-value decomposition of $M^{(m,n)}_\ell$ is computed directly using \path{scipy.linalg.svd}. For \path{qr}, a QR or LQ decomposition, implemented using \path{scipy.linalg.qr}, is first employed to reduce the SVD to a smaller square matrix. Denoting the matrix generically by $M$, with shape $m_1\times m_2$, the two cases are:

\begin{itemize}

\item If $m_1\geq m_2$, a QR decomposition $M=QR$ is performed, where $R$ has shape $m_2\times m_2$. An SVD $R=\widetilde{U}SV^\dagger$ is then performed on the smaller matrix $R$, giving $M=(Q\widetilde{U})SV^\dagger$.

\item If $m_1<m_2$, an LQ decomposition $M=LQ$ is obtained by performing a QR decomposition of $M^\dagger$, where $L$ has shape $m_1\times m_1$. An SVD $L=US\widetilde{V}^\dagger$ is then performed on the smaller matrix $L$, giving $M=US(\widetilde{V}^\dagger Q)$.

\end{itemize}

For \path{eig}, the singular values and singular vectors are obtained by diagonalizing the smaller of the two Gram matrices $MM^\dagger$ and $M^\dagger M$ using \path{scipy.linalg.eigh}. Considering again a matrix $M$ of shape $m_1\times m_2$, the two cases are:

\begin{itemize}

\item If $m_1\leq m_2$, the matrix $MM^\dagger$, of shape $m_1\times m_1$, is diagonalized as $MM^\dagger=US^2U^\dagger$. The singular values are obtained as the square roots of its eigenvalues, while the remaining singular vectors are reconstructed as $V^\dagger=S^{-1}U^\dagger M$.

\item If $m_1>m_2$, the matrix $M^\dagger M$, of shape $m_2\times m_2$, is diagonalized as $M^\dagger M=VS^2V^\dagger$. The singular values are again obtained as the square roots of its eigenvalues, while the remaining singular vectors are reconstructed as $U=MVS^{-1}$.

\end{itemize}

The \path{standard} and \path{qr} methods retain comparable numerical precision, while the latter can be faster for rectangular matrices. The \path{eig} method can provide a further computational advantage. However, because the construction of the Gram matrix squares the singular values, singular values smaller than approximately the square root of machine precision, relative to the largest singular value, may not be resolved accurately.

\end{appendix}

\bibliography{references}

\end{document}